\documentclass[twocolumn,showpacs,amsmath,prl]{revtex4-1}

\usepackage{graphicx,bm}
\usepackage{epsfig,psfrag}
\usepackage{amsmath}
\usepackage{amssymb}
\usepackage{amsbsy}
\usepackage{amsthm}
\usepackage{amsfonts}
\usepackage{wasysym}
\usepackage{bbm}
\usepackage{tabularx}
\usepackage{euscript}
\usepackage{color}
\usepackage{enumerate}
\usepackage{amsfonts}
\usepackage{exscale}
\usepackage{bbold}
\usepackage{float}

\usepackage[colorlinks,citecolor=blue]{hyperref}

\newcommand{\Fig}[1]{Fig.~\ref{#1}}

\renewcommand{\v}[1]{{\boldsymbol{#1}}}

\newcommand{\Eq}[1]{Eq.~(\ref{#1})}
\newcommand{\nn}{\nonumber\\}

\newcommand{\Rmnum}[1]{\expandafter\@slowromancap\romannumeral #1@}

\newcommand{\be}{\begin{eqnarray}}
\newcommand{\ee}{\end{eqnarray}}
\newcommand{\beq}{\begin{equation}}
\newcommand{\eeq}{\end{equation}}
\newcommand{\bpm}{\begin{pmatrix}}
\newcommand{\epm}{\end{pmatrix}}
\newcommand{\bal}{\begin{aligned}}
\newcommand{\eal}{\end{aligned}}
\newcommand{\<}{\langle}
\renewcommand{\>}{\rangle}

\newcommand{\p}{\partial}
\renewcommand{\t}[1]{{\tilde #1}}

\newcommand{\ra}{\rightarrow}

\newcommand{\e}{\epsilon}
\newcommand{\ga}{\gamma}
\newcommand{\Ga}{\Gamma}

\renewcommand{\th}{\theta}

\newcommand{\s}{{\sigma}}

\newcommand \ti[1]{}

\renewcommand{\t}{\tau}

\begin{document}

\title{Does the existence of Majorana zero mode in superconducting vortices imply the superconductivity is topologically non-trivial ?}
\author{Lokman Tsui$^{1}$, Zi-Xiang Li$^{1}$,Yen-Ta Huang$^{1}$, Steven G. Louie$^{1,2}$}
\author{Dung-Hai Lee$^{1,2}$}\email{Corresponding author: dunghai@berkeley.edu}

\affiliation{
$^1$ Department of Physics, University of California, Berkeley, CA 94720, USA.\\
$^2$ Materials Sciences Division, Lawrence Berkeley National Laboratory, Berkeley, CA 94720, USA.
}



\date{\today}

\begin{abstract}
We show that the presence of Majorana zero modes (2D), and chiral-dispersing Majorana
modes (3D), in the vortex cores of superconductors are neither sufficient nor necessary conditions for one to conclude the superconductivity is topologically non-trivial.  We discuss the relevance of this result to the proximity-induced superconductivity, in the presence of magnetic field, on the surface of topological insulators.
\end{abstract}


\maketitle

{\bf Introduction} \indent In their pioneer work\cite{ReadGreen} Read and Green showed that the  spin-polarized 2D superconductor with $p_x+ip_y$ pairing symmetry is topological. The signatures of it include the existence of gapless chiral Majorana edge states and Majorana zero modes in the cores of odd-vorticity vortices. Due to the relevance to topological quantum computing, Majorana zero modes has attracted strong interests in both theoretical and experimental communities\cite{Alicea}.

\indent In the seminal work Ref.\cite{FuKane}, Fu and Kane proposed a way to engineer a superconductor which hosts vortex Majorana zero modes. This is achieved by coupling an s-wave superconductor to the surface states of a 3D topological insulator via the proximity effect. Following this proposal, many experimental groups carry out the search of vortex Majorana zero modes.  Examples include Bi$_2$Se$_3$/NbSe$_2$\cite{JFJia} and recently Fe(Te$_{1-x}$Se$_x$)\cite{HDing}. For the latter system due to the strong spin-orbit interaction of the Te atoms, a band inversion occurs at the Brillouin zone center resulting in topologically non-trivial surface states. Ref.\cite{HDing} reports the observation of a superconducting gap on the Fermi circle associated with the surface states, which is induced by  bulk superconductivity via the proximity effect.

\indent In the literature, the existence of vortex Majorana zero modes is commonly taken as the hallmark of topologically non-trivial superconductivity. In this work, we re-examine the validity of this association. We conclude that vortex {\it Majorana zero mode is neither a sufficient nor a necessary condition for having topological superconductivity. }

 {\bf The bulk and defect classification}
\indent Our result is based on the bulk\cite{Kitaev2009,Wen2012}  and defect\cite{Teo} topological classification summarized in
table \ref{tab:realclass} of the supplementary material (SM).
In the following, we briefly explain the meaning of this table, and how to apply it to our problem.

\indent The symmetry groups given in the fourth column of table \ref{tab:realclass} are grouped into symmetry classes each labeled by either a mod 8 (the real class) or a mod 2 (the complex class)  integer $p$ in the first column of the table. For our purpose, it is sufficient to concentrate on symmetries generated by
$\hat{Q},\hat{T},\hat{C}$. They denote, respectively, the U(1) charge conservation, time reversal, and charge conjugation symmetries. Moreover we shall focus on the ``real class'' of the classification table\cite{Kitaev2009,Wen2012}.

\indent The generic Hamiltonian classified by table \ref{tab:realclass} has the form
\be
H=\int d^dx~\chi^T(\v x)\left(-i\sum_{j=1}^d \gamma_j{\p_j}+i \sum_{\alpha=1}^m \phi_\alpha M_\alpha\right)\chi(\v x).\nn&&\label{H}\ee Here $\chi(\v x)$ is a $q$-component Majorana fermion field, where $q$ is the number of spin and orbital degrees of freedom in the problem. In the following we shall refer to the internal space spanned by the spin and orbital degrees of freedom as the ``mode space''\cite{Kitaev2009}.  The $\gamma_j$ are $q\times q$ real symmetric  matrices satisfying $\{\gamma_j,\gamma_l\}=2\delta_{jl}$.  The ``mass matrices'' $M_\alpha$ are $q\times q$ real antisymmetric matrices satisfying $\{\gamma_j,M_\alpha\}=0$. The $Q,T,C$ in table \ref{tab:realclass} are $q\times q$ real orthogonal matrices. They are the representations of $\hat{Q},\hat{T},\hat{C}$ in the mode space. Both $\gamma_j$ and $M_\alpha$  anti-commute with $T$ but commute with $Q$ and $C$.

\indent The $\{\phi_\alpha~|\alpha=1,...,m\}$ in \Eq{H} are real scalars, their presence opens the fermion gap hence they play the role of order parameters. The ``mass manifolds'' are manifolds in  the $R^m$ space spanned by $\{\phi_\alpha\}$. If  $\{\phi_\alpha\}\in$ a mass manifold, the fermion gap stays a constant. Thus mass manifold has the same meaning as  ``target space'' for non-linear sigma models. Fixing the fermion gap to, say, unity,  the mass manifolds are the solutions of $(\sum_{\alpha=1}^m \phi_\alpha M_\alpha)^2=-I_{q\times q}$ (the identity $q\times q$ matrix).

\indent The union of all mass manifolds is the ``classifying space''. Because each mass matrix is required to anti-commute with all gamma matrices $\{\gamma_j, j=1,...,d\}$, for a fixed $q$ the number of $M_\alpha$ (i.e. $m$) depends on the spatial dimension $d$. As a result, both the mass manifolds and the classifying space depend on the spatial dimension. The integer $n$  is equal to $q/q_0$ where $q_0$ is the smallest mode space dimension necessary to represent the particular symmetry class in $d=0$.
The $R_p$ listed in the second column of \ref{tab:realclass} are the associated classfying space.

\indent The abelian groups  given in the third column of  table \ref{tab:realclass} are  the zero$^{\rm th}$ homotopy group of the classifying space $R_p$, i.e., $\pi_0(R_p)$. The number of elements in such group is equal to the number of disconnected mass manifolds, each of which corresponds to an equivalent class of topological phases. In general  $\pi_0(R_p)$ depends on $n$,  what listed is the ``stabilized'' result, namely  $\pi_0(R_p)$  in the limit $n$ tends to infinity. (Note that as $n\ra\infty$ so do $q$ and $m$.)

\indent To find the bulk classification one first locates the appropriate symmetry class $p$. For example, if the problem has  time reversal ($\hat{T}^2=-1$) and charge conservation ($\hat{Q}$) symmetries we denote the symmetry group as $ G_-(Q,T)$ (see the caption of table \ref{tab:realclass} for the explanation of the symbol). The symmetry class is $p=4$. Next, we find the classifying space for a given space dimension $d$. The result (proof not given) is $R_{p-d}$ (for $d=0$ it is $R_p$ as expected). Thus for the symmetry group $ G_-(Q,T)$  and $d=3$ the classifying space is $R_1$, for which the  bulk classification is $\pi_0(R_1)=Z_2$. This means there are two inequivalent classes of fully gapped free fermion
phases. Phases within the same class can go into each other without closing the energy gap or breaking the symmetries. These classes are represented by the elements of $Z_2$:  $0$ stands for the trivial class (phases in this class do not have gapless boundaries) and $1$ stands for the non-trivial class (phases in this class possess gapless boundary excitations). The group operation of $Z_2$ corresponds to the ``stacking'' of different phases on top of each other. This means tensoring the mode spaces and turning on all symmetry allowed local interactions between the corresponding spin and orbital degrees of freedom. For example, $1+1=0$ means when two non-trivial phases are stacked together the result is a trivial phase, a phase with no gapless boundary.
\begin{figure}
\includegraphics[scale=0.2]{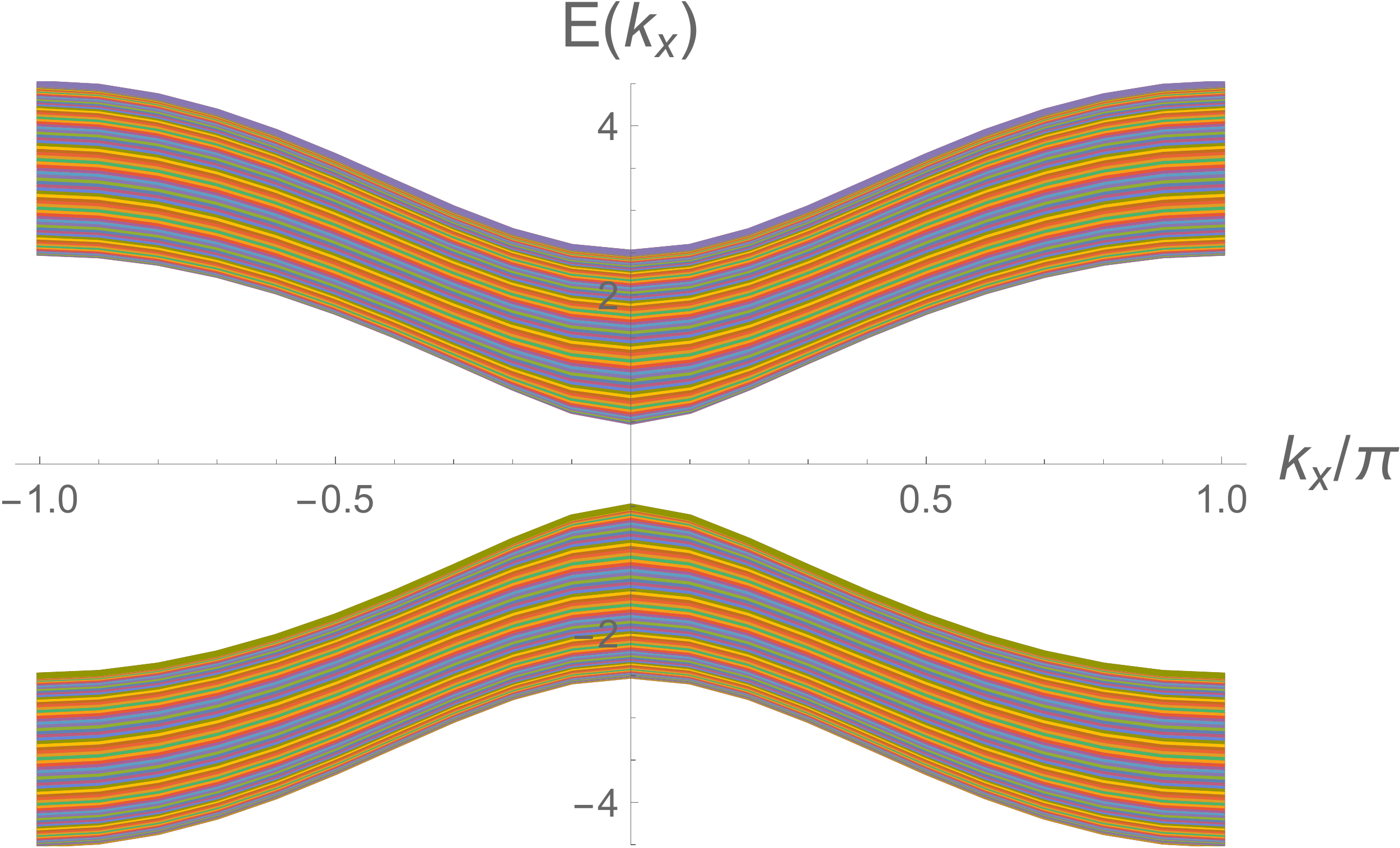}
\caption{Projected band structure of \Eq{lathx} with periodic boundary condition in $ \hat{x} $ and open boundary condition in $ \hat{y} $. The number of rows in y direction, $n_y$, is $80$. The $\phi_{1,2}$ used to generate this plot are $\phi_1=0.5\cos\theta$ and $\phi_2=0.5\sin\theta$ we have checked that for any angle (the figure shows when $\theta=\pi/8$ 
there is no gapless edge mode signifying the bulk are trivial superconductors.}
\label{fig1}
\end{figure}

\indent Defects are singularities in the free fermion Hamiltonian. For example, a vortex is a point defect in the BdG Hamiltonian of a superconductor. It is important to remember that in finding out the defect classification \cite{Teo} one needs to specify the Hamiltonian symmetry in the presence of defects. Each defect is characterized by its codimension, the difference between the space dimension and the defect dimension, $k$. For example a point defect has $k=d$ and a line defect has $k=d-1$ ... etc.  Having determined the bulk classifying space $R_{p-d}$, the defect classification is given by $\pi_{k-1}(R_{p-d})$. According to Bott's periodicity theorem\cite{Bott1959} $\pi_{k-1}(R_{p-d})=\pi_0(R_{p-d+k-1})$.Hence the defect classification is given by the third column of the $(p-d+k-1)^{\rm th}$ row in table \ref{tab:realclass}.
For example for 2D superconductors with $\hat{T}^2=+1$  symmetry (symmetry group $G_+(T)$) the  symmetry class is $p=1$. The classification of vortices ($k=2$) is given by the third column of the $1-2+2-1=0^{\rm th}$ row in table \ref{tab:realclass}, namely, $Z$. This means a vorticity $+1$ vortex will have one Majorana zero mode and $+2$ vortex (which can be viewed as two  $v=+1$ vortices stacked together) has two Majorana zero modes ... etc. Of course, when a $+1$ vortex is stacked with a $-1$ vortex there exists symmetry allowed interaction that can gap out both zero modes.

\indent Armed with table \ref{tab:realclass} we conclude that 2D superconductor with $\hat{T}^2=+1$ symmetry (i.e. $p=1$) is trivial because $\pi_0(R_{1-2}=R_7)=0$. However its vortex classification is $\pi_0(R_{1-2+2-1}=R_0)=Z$ which is non-trivial, i.e., a vorticity $+n$ vortex will harbor $n$ Majorana zero modes. Similarly 3D superconductors with no symmetry (i.e. $p=2$) are topologically trivial because $\pi_0(R_{2-3}=R_7)=0$. However their vortex lines are classified by $\pi_0(R_{2-3+2-1}=R_0)=Z$, i.e., a vorticity $n$ vortex line harbors $n$ branches of chiral-dispersing Majorana modes in the vortex core.
\indent This proves that there can exist superconductors which have topologically trivial bulk, but in the cores of their vortices there are Majorana zero modes (2D) or branches of chiral-dispersing Majorana modes (3D).
However since the classification results of table \ref{tab:realclass} are meant for the limit where the $n (\mathrm{hence~} q)\ra \infty$, it is important to check whether the statements made in the last paragraph hold for finite $q$, i.e., for systems that might exist in experiments. To do so we need to construct explicit models with a finite number of spin and orbital degrees of freedom.

{\bf An example in 2D}
\indent Let's focus on the symmetry class $p=1$. Because we are interested in superconductors (where $\hat{Q}$ is not a good symmetry) there are only two symmetry groups: 1) $G_+(T)$ ($ T^2=+1 $ only), and 2) $G_{++}^+(T,C)$ $ (T^2=+1 $, $ C^2=+1 $ ,$ [T,C]=0 $). For symmetry group (2), $ C $ commutes with the Hamiltonian  and $ T $, so it can be first diagonalized. After doing so the problem decouples into two sectors with $ \pm 1 $ eigenvalue for $ C $. Therefore we can proceed to analyze  each eigen sector of $ C $ and the problem reduces to symmetry group (1), i.e., $ T^2=+1 $ only.

\indent For symmetry class $p=1$ in two dimensions it can be shown that $q_0=4$, and the time reversal matrix is given by $T=\e\e$. (Here we abbreviate the Kronecker product of two matrices each equal to $\e=i\s_y$ as $\e\e$.) Hence a $4\times 4$ Hamiltonian can be written down. In continuum space such a Hamiltonian read
\be
H&&=\int d^2x~\chi^T(\v x)(-i IZ~{\p_1}-i IX~{\p_2}-i\phi_1 X\e\nn&&-i\phi_2 Z\e)\chi(\v x).\label{H2d}\ee
A lattice version of this Hamiltonian is given by \Eq{lathx} of the SM. In addition, we also express \Eq{H2d} in terms of complex fermion operators (SM \Eq{cH}) so that it is easier to associate it with the BdG Hamiltonian of a superconductor.

\indent Requiring  $(\phi_1M_1+\phi_2 M_2)^2=-1 $ implies $ \phi_1^2+\phi_2^2=1 $. Therefore the mass manifold is a circle.
This is consistent with the classification table since the classifying space is $R_{1-2}=R_7=U(n)/O(n)$ which is $S^1$ for $n=4/q_0=4/4=1$. Since the classifying space consists of only one disconnected mass manifold the classification is trivial, i.e., there are only topological trivial phases.

\indent To check whether the  bulk superconductor given by \Eq{H2d} or \Eq{cH} is indeed topologically trivial we diagonalize  \Eq{lathx} in the SM on a cylinder with periodic (open) boundary condition in the $ x $ ($ y $) directions. The result is shown in \Fig{fig1}(a) which shows there is no gapless edge state, consistent with the statement that the bulk is trivial.

\begin{figure}
\includegraphics[scale=0.21]{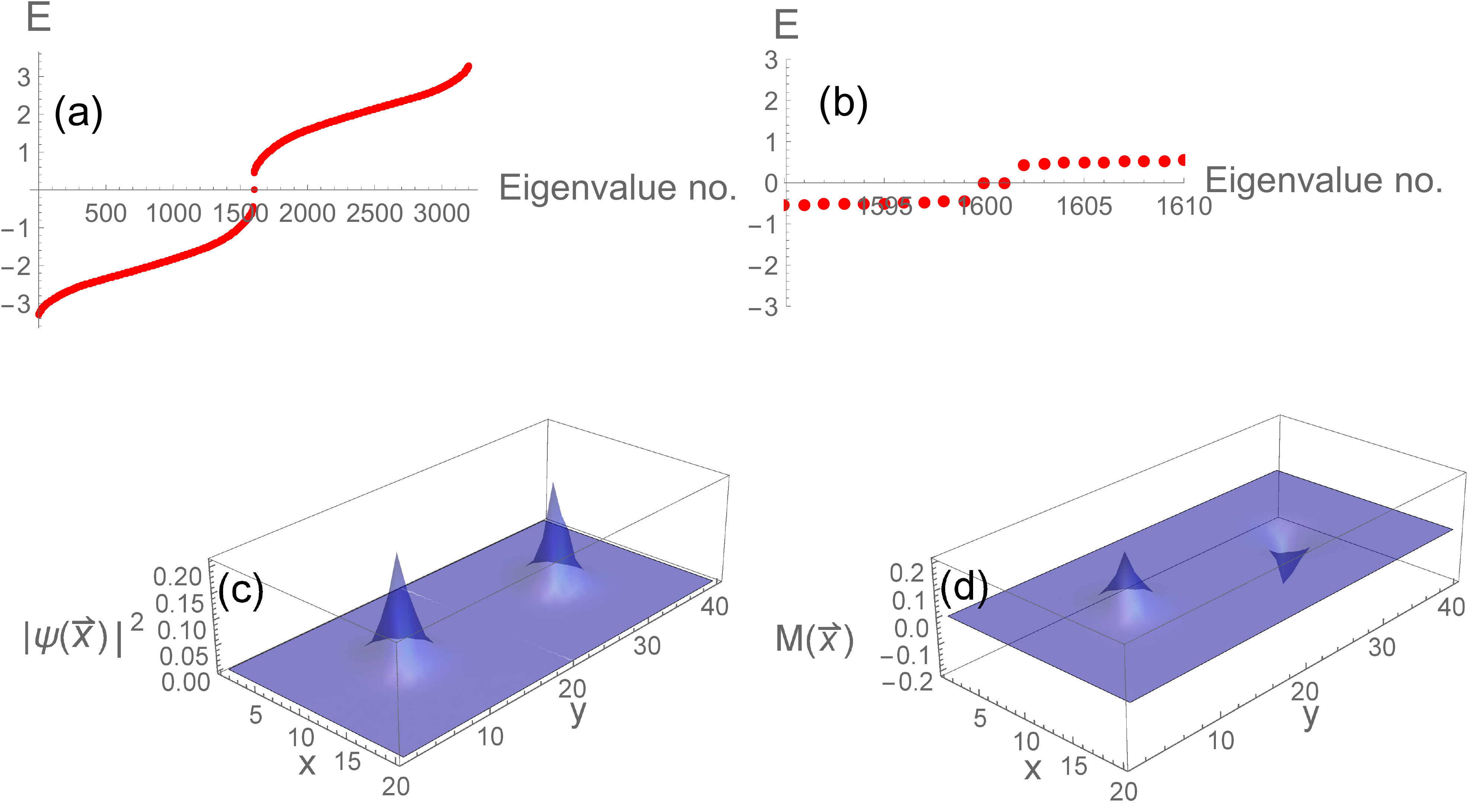}
\caption{(a) and (b) The energy spectra of \Eq{vlathx} in the presence of a pair of $\pm 1$  vortices where $\phi_{1,2}(\v x_i)$ are given by \Eq{vor} with $\Delta_0=0.5$ and $n_v=1$. Panel(b) is a zoom-in of panel (a) near zero energy. Panel (c) is the sum of the modulus square of the two zero mode eigenfunctions. Panel (d) $M(\v x)$ is defined as $\<\Phi_1(\v x)|YY|\Phi_1(\v x)\>+\<\Phi_2(\v x)|YY|\Phi_2(\v x)\>$ where $|\Phi_1(\v x)\>$ and $|\Phi_2(\v x)\>$ are the 4-component eigenfunctions of the two Majorana zero modes. The  calculation is done under open boundary condition with $20\times 40$ sites ($40/20$ in the direction parallel/perpendicular  to the separation vector between the vortices). }
\label{fig2}
\end{figure}

\begin{figure}
\includegraphics[scale=0.25]{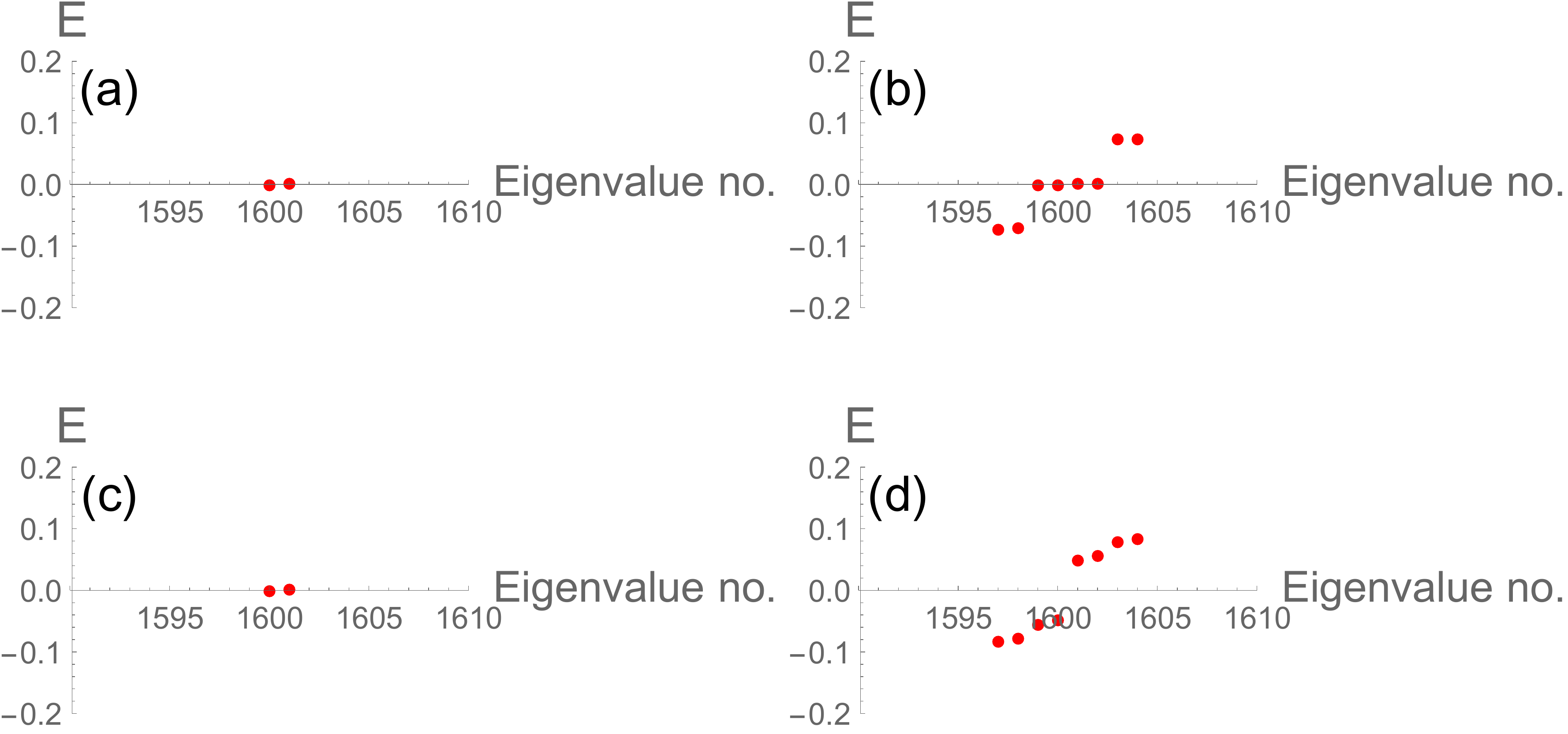}
\caption{The zoom-in (around $E=0$) energy spectra of \Eq{vlathx} with $\pm 1$ (panels (a),(c)) and $\pm 2$ (panels (b),(d)) vortices. The $\Delta_0$ used to construct these figures is equal to $.5$. Panel (a) and (b) are for $\phi_3=0$ superconductor, while panel (c) and (d) are for $\phi_3=0.2$. The system size is $20\times 40$ sites ($40/20$ in the direction parallel/perpendicular  to the separation vector between the vortices).}
\label{fig3}
\end{figure}

\indent To check whether the vortices harbor Majorana zero modes we use the following order parameters
\be
&&\phi_1(\v x)+i\phi_2(\v x)\nn&&=\Delta_0 \left[{(x-x_0)+i(y-y_0)\over |(x-x_0)+i(y-y_0)|}{(x-x_1)-i(y-y_1)\over |(x-x_1)-i(y-y_1)|}\right]^{n_v},\nn&&
\label{vor}\ee
where $(x_0,y_0)$  and $(x_1,y_1)$ are the the centers of a pair of $\pm n_v$ vortices. In \Fig{fig2}(a) and (b) we show the eigen spectra of \Eq{vlathx} for the case where $n_v=1$. \Fig{fig2}(b) is a zoom-in of \Fig{fig2}(a) near the zero eigenvalue. It shows that there are a pair of zero modes, one associated with the vortex and the other associated with antivortex, respectively. This result is supplemented by the analytic solution of the Majorana zero mode in a single vortex in the SM. In \Fig{fig2}(c) we plot the sum of the modulus square of the eigen functions associated with the pair of zero modes. It clearly shows that the zero modes are localized in the core of the vortices.  According to the analytic solution, the wavefunctions of the Majorana zero modes should be the eigenfunctions of $YY$ with eigenvalue $\pm 1$ ($+$ for the vortex and $-$ for the anti-vortex). In \Fig{fig2}(d) we plot the the sum of the expectation value of $YY$ in the vortex/antivortex zero modes.  The result is consistent with the analytic solution.
In addition to the above  results, in \Fig{fig3}(b) we have also studied the energy spectrum of a pair of $\pm 2$ vortices. Compared with that of the $\pm 1$ pairs (\Fig{fig3}(a)) the number of zero modes has doubled. This is consistent with the classification of the vortex being $Z$.

{\bf An example  in 3D} \indent With a slight generalization we can write down the Hamiltonian for the topologically trivial,
symmetry class $p=2$ (no symmetry), superconductors in $3D$. The only change is to add one more gamma matrix to \Eq{H2d}, i.e.,
\begin{figure}
\includegraphics[scale=0.2]{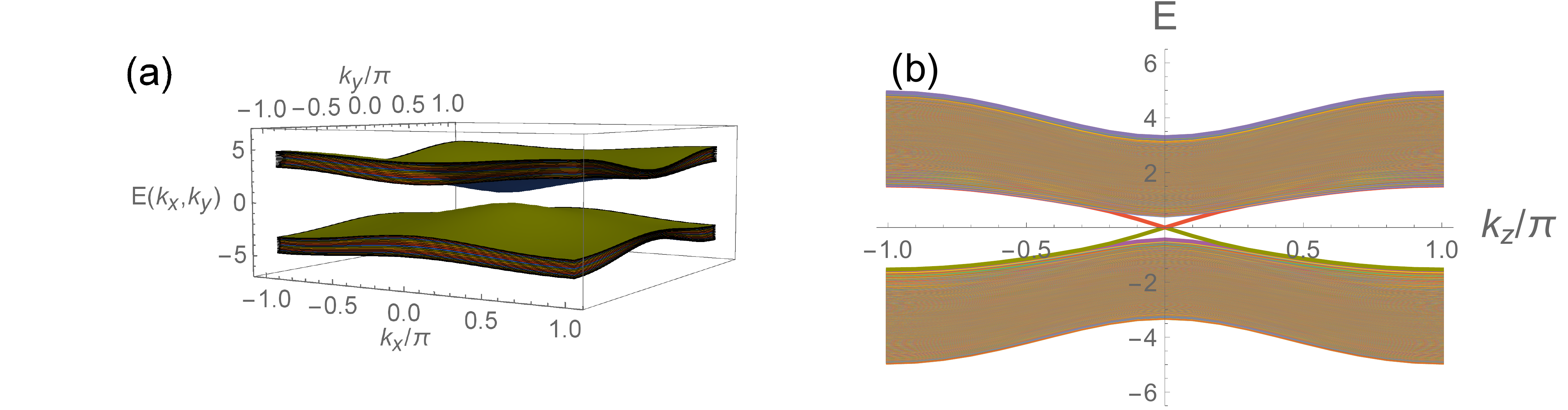}
\caption{(a) The projected bandstructure of \Eq{lathx3d}  with periodic boundary condition in $ \hat{x} $, $\hat{y}$ and open boundary condition in $ \hat{z} $. There are $n_z=40$ rows in the $\hat{z}$ direction. The $\phi_{1,2}$ used to generate this plot are $\phi_1=0.5 \cos(\pi/8)$ and $\phi_2=0.5\sin(\pi/8)$. (b) The spectrum of \Eq{vlathx3d} in the presence of a pair of $\pm 1$ vortices. The vortex lines are running in the $\hat{z}$ direction which is subjected to periodic boundary condition.
The boundary condition in $\hat{x}$ and $\hat{y}$ are open. The number of sites parallel/perpendicular to the vector separating the two vortices are $40/20$, respectively. The $\Delta_0$ in \Eq{vor} is set to 0.5. }
\label{fig4}
\end{figure}
\be
H&&=\int d^3x~\chi^T(\v x)(-i IZ~{\p_1}-i IX~{\p_2}-i\e\e~\p_3\nn&&-i\phi_1 X\e-i\phi_2 Z\e)\chi(\v x).\label{H3d}\ee
The lattice version of this Hamiltonian is given by \Eq{lathx3d} of the SM.
In \Fig{fig4}(a) we plot the projected bandstructure for a system periodic in $\hat{x}$ and $\hat{y}$ and open in $\hat{z}$. The absence of gapless boundary modes signifies the topological trivialness of the bulk superconductor. In \Fig{fig4}(b) we plot the energy spectrum of a pair of $\pm 1$ vortices with their core line running in the $z$ direction. Here the
boundary conditions are open in $\hat{x},\hat{y}$ and periodic in $\hat{z}$. A pair of chiral-dispersing gapless Majorana modes are present, with the up/down moving branch localized on the vortex/antivortex, respectively.  Again, we have a trivial superconductor harboring  non-trivial chiral Majorana modes in the vortex cores !

{\bf Experimental relevance} \indent  The surface Dirac cone of a 3D topological insulator (such as Bi$_2$Se$_3$) is protected by the $\hat{T}^2=-1$ and the $\hat{Q}$ symmetries. After the proximity effect induces superconductivity and the application of magnetic field (to induce vortices) the $\hat{T}$ symmetry is also broken. Under such condition the symmetry class is $p=2$ (no symmetry). The topological classification of the $p=2$ symmetry class is $Z$. For the $q=4$ (which applies to the surface states of 3D topological insulators) there are three topologically inequivalent phases. The mass manifolds consist of two points (which correspond to quantized anomalous Hall phases\cite{yu,xue} with opposite $\s_{xy}$), and a circle (which correspond to a superconductor). In \Fig{gapclosure}(a) of the SM we show these order parameter manifolds (red) and the surface (blue) at which the fermion gap closes. The hamiltonian describing these three phases is given by
\be
H&&=\int d^2x~\chi^T(\v x)(-i IZ~{\p_1}-i IX~{\p_2}-i\phi_1(\v x) X\e\nn&&-i\phi_2(\v x) Z\e-i\phi_3 I\e)\chi(\v x).\label{H2dmu}\ee
The hamiltonian associated with the circle  in \Fig{gapclosure}(a) is identical to that in \Eq{H2d}. The extra $T^2=+1$ symmetry is broken
by shifting the circle up as shown in \Fig{gapclosure}(b). Because such shift does not cross the gap closing surface, the hamiltonians are adiabatically connected hence describing the same topological phase. However due to the breaking of the $T^2=+1$ symmetry the vortex classification become $Z_2$, which agrees with the classification of the vortices of the $p=2$ symmetry class $\pi_0(R_{2-2+1})=Z_2$.\\

In \Fig{fig3} we compare the vortex energy spectra for the superconductors associated with the circles in \Fig{gapclosure}(a,b). \Fig{fig3}(a) and (b) are the zoom-in energy spectra for (a) $\pm 1$ and (b) $\pm 2$ vortex pairs in the $T^2=+1$ superconductor. The spectra in \Fig{fig3}(c) and (d) are for the vortices of the no-symmetry superconductor. While the $\pm 1$ vortices still possess a Majorana zero mode, there is no zero mode for the $\pm 2$ vortices!  This is consistent with the vortex classification being $Z_2$\cite{FuKane}.
In addition, in the SM we show that the interface between the quantized anomalous Hall and the superconducting phases possesses chiral Majorana modes\cite{Qi} consistent with these phases being topologically inequivalent.  These results suggest that in the presence of the magnetic field, the proximity-induced superconductivity on the surface of 3D topological insulator is topologically trivial.\\

{\bf Discussions and conclusion}\indent So far we have shown that the presence of vortex Majorana modes is not a sufficient  condition  for concluding the parent superconductor is topologically non-trivial. However is it a necessary condition? The answer is negative. The superconductor with symmetry $G_{--}^-(T,C)$ in the symmetry class $p=4$ is an example. In 2D the classifying space is given by $R_{4-2}=R_2$. Its classification ($\pi_0(R_2)$) is $Z_2$ hence is non-trivial. However the vortices are classified by $\pi_0(R_{2+2-1})=\pi_0(R_3)=0$.

\indent The above results lead us to conclude that the presence of vortex Majorana zero mode (2D) or chiral-dispersing Majorana modes (3D)  in the vortex core of superconductors
are neither sufficient nor necessary condition for concluding the superconductivity being topological. Of course, this conclusion does not affect the novelty of the Majorana modes and their possible applications.\\

{\bf Acknowledgement}\indent We thank Prof. Fa Wang for useful discussions. This work was supported by the Theory Program at the Lawrence Berkeley National Laboratory, which is funded by the U.S. Department of Energy, Office of Science, Basic Energy Sciences, Materials Sciences and Engineering Division under Contract No. DE-AC02-05CH11231

\newpage
\begin{widetext}
\section{Supplementary Material}
\section{Table I}
\begin{table*}[h]
\centering
\caption{The classification tables. Here $G$ denotes the symmetry group and  $\hat{Q},\hat{T},\hat{C} $ are generators of the
U(1) charge conservation, time reversal and charge conjugation symmetries, respectively. $ Q,T,C $ are orthogonal matrices representing  these generators in the mode space, i.e. the spin and orbital space of Majorana fermion operators $\chi_a(\v x)$. More specifically, 
$ \hat{Q}= \sum_{a,b} \frac{i}{4}\chi_a(\v x) Q_{ab}\chi_b(\v x)$, $\hat{T}~i~\hat{T}^{-1}=-i$ and $\hat{T}\chi_a(\v x) \hat{T}^{-1}=\sum_bT_{ab}\chi_b(\v x)$, $\hat{C}\chi_a(\v x) \hat{C}^{-1}=\sum_bC_{ab}\chi_b(\v x)$.  The meaning of the signs in the expressions, e.g., $ G_{S_T}(T) $, $ G_{S_C}(C) $, $ G^{S_{TC}}_{S_T S_C}(T,C) $ and $ G^{S_{TC}}_{S_T S_C}(Q,T,C) $ are the following: $ T^2=(-1)^{S_T} $, $ C^2=(-1)^{S_C} $, and $ TC=(-1)^{S_{TC}} CT $. In this table it is assumed that $ Q^2=-1$ and $TQ=-QT $ hence $S_Q$ and $S_{TQ}$ are not specified.
}
\vspace{0.1 in}
\center{Real class}\\
\vspace{0.1in}
\begin{tabular}{ |c|c|c|c|}
\hline
$ p $ mod 8&Classifying spaces $ R_p $ & $ \pi_0 (R_p) $ & Symmetry generators\\
\hline
0&$\bigcup \frac{O(n)}{O(l)\times O(n-l)}$&$Z$&$ G_+(Q,T),G^{-}_{+-}(T,C) $\\
\hline
1&$O(n)$&$Z_2$&$ G_+(T),G^{+}_{++}(T,C),G^{\pm}_{++}(Q,T,C) $\\
\hline
2&$\frac{O(2n)}{U(n)}$&$Z_2$&$none, G_+(C),G^{-}_{\pm+}(T,C),G_{+}(Q,C) $\\
\hline
3&$\frac{U(2n)}{Sp(n)}$&$0$&$ G_-(T),G^{+}_{-+}(T,C),G^{\pm}_{-+}(Q,T,C) $\\
\hline
4&$\bigcup \frac{Sp(n)}{Sp(l)\times Sp(n-l)}$&$Z$&$ G_-(Q,T),G^{-}_{--}(T,C) $\\
\hline
5&$Sp(n)$&$0$&$ G^{\pm}_{--}(Q,T,C)$\\
\hline
6&$\frac{Sp(n)}{U(n)}$&$0$&$ G_-(Q,C),SU(2) $\\
\hline
7&$\frac{U(n)}{O(n)}$&$0$&$ G^{\pm}_{+-}(Q,T,C)$\\
\hline
\end{tabular}
\label{tab:realclass}
\centering
\vspace{0.3 in}
\center{Complex class}\\
\vspace{0.1in}
\begin{tabular}{ |c|c|c|c|}
\hline
$ p $ mod 2&Classifying spaces $ C_p $ & $ \pi_0 (C_p) $ & Symmetry generators\\
\hline
0&$\bigcup \frac{U(n)}{U(l)\times U(n-l)}$&$Z$&$ U(1),G_{-}(C) $\\
\hline
1&$U(n)$&$0$&$G^{+}_{--}(T,C),G^{+}_{+-}(T,C)$\\
\hline
\end{tabular}
\label{tab:cpxclass}
\end{table*}

\section{The complex fermion version of \Eq{H2d}}
Here we recast \Eq{H2d} into complex fermion form $ \psi(\v x)=(\chi_{1}(\v x)-i\chi_{2}(\v x))/2$ where $1$ and $2$ are the indices of the first Pauli matrix in \Eq{H2d}:
\be
H&&=\int d^2 x~4\psi^{\dagger}(-i Z\p_1 -i X\p_2 )\psi(\v x)+2\Big[ (\phi_1(\v x)-i\phi_2(\v x))\nn&&\psi^{T}(\v x)~\e ~\psi(\v x)+ h.c.\Big].
\label{cH}\ee

\section{The 2D lattice version of \Eq{H2d}}

Here we write down a Hamiltonian on a square lattice which has the low energy effective theory described by \Eq{H2d}. We first do so in the momentum space.

\begin{align}
H=\sum_{\v k\in BZ} \chi_{-\v k}^{T} \left[\sin(k_x) IZ+\sin(k_y) IX- (1-\cos(k_x))iX\e - (1-\cos(k_y))iZ\e
 -\phi_1 iX\e - \phi_2 iZ\e \right]\chi_\v k \label{lathk}
\end{align}

In real space, the Hamiltonian is
\begin{align}
H=\sum_i -\chi_{i}^{T} \left[(1+\phi_1)iX\e + (1+\phi_2) iZ\e \right]\chi_i
 + \sum_i \left[\chi_{i}^{T} \left(\frac{-i IZ+iX\e}{2}\right)\chi_{i+\hat{x}}+\chi_{i}^{T} \left(\frac{-i IX+iZ\e}{2}\right)\chi_{i+\hat{y}} + h.c.\right] .\label{lathx}
\end{align}


\section{The 2D lattice Hamiltonian in the presence of vortices}
In the presence of vortices the lattice Hamiltonian we diagonalized to obtain \Fig{fig2} is given by
 \be
H&&=\sum_i -\chi_{i}^{T} \left[(1+\phi_1(\v x_i))iX\e + (1+\phi_2(\v x_i)) iZ\e +\phi_3 i I\e\right]\chi_i
 + \sum_i \Big[\chi_{i}^{T} \left(\frac{-i IZ+iX\e}{2}\right)\chi_{i+\hat{x}}+\chi_{i}^{T} \left(\frac{-i IX+iZ\e}{2}\right)\chi_{i+\hat{y}}\nn&& + h.c.\Big] ,\label{vlathx}\ee where $\phi_{1,2}(\v x_i)$ are given by \Eq{vor}.

\section{Analytic solution of the Majorana zero mode in a  vorticity +1 vortex}
The low energy effective Hamiltonian is
\begin{align}
H=\int d^2x~\chi^{T}(\v x) (-i IZ\p_1-i IX\p_2-i\phi_1(\v x)X\e-i\phi_2(\v x)Z\e) \chi(\v x) \label{eqn:H0}
\end{align}
where \be
\phi_1(\v x)+i\phi_2(\v x)=\Delta_0 \left({x+iy\over |x+iy|}\right)^{n_v}.\label{vor1}\ee
By going to the polar coordinate the Hamiltonian becomes
\begin{align}
H&=\int r dr d\th~\chi^{T}(r,\th) \left[-i IZ\left(\cos\th\p_r-{\sin\th\over r}\p_\th\right)-i IX\left(\sin\th\p_r+{\cos\th\over r}\p_\th\right)-i(\phi_1(\v x) X\e+\phi_2(\v x) Z\e)\right] \chi(r,\th) \nn
&=\int r dr d\th~\chi^{T}(r,\th)  \left[e^{-i\th}(IZ+iIX)/2\left(-i\p_r-{1\over r}\p_\th\right)+\Delta_0 e^{-in_v\th}(XY+iZY)/2+h.c.\right]\chi(r,\th)
\label{eqn:H0}
\end{align}

It is straightforward to check that $H$ is invariant under rotation by angle $\phi$, namely $R(\phi)HR^{-1}(\phi)=H$, where
$ R(\phi)=e^{i\hat{J}\phi} $ and $ \hat{J}=\frac{1}{i}\p_{\th}-\frac{1}{2}(n_v YI-IY) $. The two terms are the orbital and spin angular momentum, respectively.

The eigenfunctions of $ \hat{J} $ has the form $ e_{\rho \s}e^{im\th} $ with eigenvalues $ m-\frac{1}{2}(n_v\rho-\s) $. Here $ e_{\rho \s} $ is a four-component spinor with eigenvalues $ \rho,\s=\pm1 $ under $ YI,IY $ respectively. Their explicit forms are
\begin{align*}
e_{++}^T&=
\frac{1}{2}(1,i,i,-1)\\
e_{+-}^T&=
\frac{1}{2}(1,-i,i,1)\\
e_{-+}^T&=
\frac{1}{2}(1,i,-i,1)\\
e_{--}^T&=
\frac{1}{2}(1,-i,-i,-1)
\end{align*}

We expand $ \chi(r.\th) $ in eigenfunctions of $ \hat{J} $ as
\begin{align*}
\chi_{\alpha}(r,\th) =\sum (e_{\s \t})_{\alpha} e^{im\th} \ga_{\s\t m}(r)
\end{align*}

and substitute into \eqref{eqn:H0}, the result is
\be
H&&=\int 2\pi rdr \sum_{l\in \mathbb{Z}} \Gamma_j^{\dagger}(r) \bpm
0 & -i\p_r-il/r & -i\Delta_0 & 0 \\
-i\p_r+i(l-1)/r & 0 & 0 & i\Delta_0 \\
i\Delta_0 & 0 & 0 & -i\p_r-i(l-n_v)/r \\
0 & -i\Delta_0 & -i\p_r+i(l-n_v-1)/r & 0
 \epm \Gamma_j(r) \nn&&
=\int 2\pi r dr \sum_j \Gamma_j^{\dagger}(r)D_j\Gamma_j(r)
\ee
where $ l=j+\frac{1+n_v}{2} $ and
\begin{align*}
D_j&=IX(-i\p_r-i/2r)+IY(j/r)+ZY(n_v/2r)+YZ\Delta_0 \\
 \Gamma_j^T&=(\ga_{++(l-1)},\ga_{+-l},\ga_{-+(l-n_v-1)},\ga_{--(l-n_v)})
\end{align*}
Note that $ \ga $'s are complex fermions operators. Since $ \chi^\dagger=\chi $ and $ e_{\s\t}^*=e_{-\s-\t} $, we deduce $ \ga_{\s\t m}^\dagger=\gamma_{-\s-\t-m} $. Therefore $ \Gamma_j^{*}=XX\Gamma_{-j} $. 

Also a useful identity is
\begin{align}
\int 2\pi rdr \sum_{\alpha}f_{\alpha}(r)[D_j g(r)]_{\alpha}= \int 2\pi rdr \sum_{\alpha} \left[-XXD_{-j}XX f(r)\right]_{\alpha} g_{\alpha}(r)\label{eqn:identity}
\end{align}
from which it can be shown that $ \int 2\pi r dr \Gamma_j^{\dagger}(r)D_j\Gamma_j(r)=\int 2\pi r dr \Gamma_{-j}^{\dagger}(r)D_{-j}\Gamma_{-j}(r)$.

It turns out a normalizable zero mode solution exists\cite{JackiwRossi} when $ n_v=1 $ and $ j=0 $. The zero mode solution $ \xi=\int 2\pi rdr f_{\alpha}(r)[\Ga_0^{\dagger}(r)]_{\alpha} $ satisfies $ [H,\xi]=0 $. By using \eqref{eqn:identity} and $ \{[\Ga_j^*(r)]_{\alpha},[\Ga_{j'}(r')]_{\beta}\}=\delta_{\alpha\beta}\delta_{j+j'}\frac{1}{\pi r}\delta(r-r') $,
this implies $ D_0 f=0 $. The zero mode solution is $ f=e^{-\Delta_0 r} (i,0,0,-i) $. The phase is chosen such that $ \xi^{*}=\xi $.

Note the $ YY $ in the $ \chi $ basis is $ \text{diag}(1,-1,-1,1)=ZZ $ in the $ \Gamma $ basis. It is also readily seen that $ (ZZ)f=+f $.

For $ n_v=-1 $, the zero mode solution is $ f=e^{-\Delta_0 r} (0,1,1,0) $. It is seen that $ (ZZ)f=-f $.


\section{The 3D lattice Hamiltonian}
The 3D version of \Eq{lathk} is given by
\begin{align}
H&=\sum_{\v k\in BZ} \chi_{-\v k}^{T} \Big[\sin(k_x) IZ+\sin(k_y) IX+\sin(k_z) \e\e- (1-\cos(k_x))iX\e - (1-\cos(k_y))iZ\e - (1-\cos(k_z))iZ\e \nn
  &-\phi_1 iX\e - \phi_2 iZ\e \Big]\chi_\v k \label{lathk3d}
\end{align}

The 3D version of \Eq{lathx} is given by
 \be
H&&=\sum_i -\chi_{i}^{T} \left[(1+\phi_1)iX\e + (2+\phi_2) iZ\e \right]\chi_i
 + \sum_i \Big[\chi_{i}^{T} \left(\frac{-i IZ+iX\e}{2}\right)\chi_{i+\hat{x}}+\chi_{i}^{T} \left(\frac{-i IX+iZ\e}{2}\right)\chi_{i+\hat{y}}\nn&& + \chi_{i}^{T} \left(\frac{-i \e\e+iZ\e}{2}\right)\chi_{i+\hat{z}}+h.c.\Big] ,\label{lathx3d}\ee

 and the 3D version of \Eq{vlathx} is given by
 \be
H&&=\sum_i -\chi_{i}^{T} \left[(1+\phi_1(\v x_i))iX\e + (2+\phi_2(\v x_i)) iZ\e \right]\chi_i
 + \sum_i \Big[\chi_{i}^{T} \left(\frac{-i IZ+iX\e}{2}\right)\chi_{i+\hat{x}}+\chi_{i}^{T} \left(\frac{-i IX+iZ\e}{2}\right)\chi_{i+\hat{y}}\nn&& + \chi_{i}^{T} \left(\frac{-i \e\e+iZ\e}{2}\right)\chi_{i+\hat{z}}+h.c.\Big] ,\label{vlathx3d}\ee
where $\phi_{1,2}(\v x_i)$ are given by \Eq{vor}.

\section{Chiral Majorana modes at the interface between superconductivity and ferromagnetism }
In the following we will show that in the interface between superconductivity and ferromagnetism, there is a chiral majorana mode. We show this by two methods: first by analytically solving the low energy effective Hamiltonian in continuum, and second by numerically diagonalize the corresponding lattice Hamiltonian.

\subsection{Analytical solution}
The low energy effective Hamiltonians for the 2D superconductivity and quantized anomalous Hall phases are given respectively by:
\begin{align}
H_{SC}&=\int d^2x~\chi^T(\v x)(-i IZ~{\p_1}-i IX~{\p_2}-i\phi_1 X\e-i\phi_2 Z\e)\chi(\v x)\nn
H_{FM}&=\int d^2x~\chi^T(\v x)(-i IZ~{\p_1}-i IX~{\p_2}-i\phi_3 I\e)\chi(\v x)\label{eq:scfm}
\end{align}
Assume a configuration where for $ x<0 $, the system is in the ferromagnetic phase with uniform $\phi_3$, whereas for $ x\geq0 $, the system is in the superconducting phase with uniform $ \phi_1,\phi_2 $. Furthermore, we may perform a local orthogonal transformation without affecting the kinetic terms and the ferromagnetic term, such that $ \phi_1=0 $ and $ \phi_2=\phi>0 $. So
\begin{align*}
H=&\begin{cases}
\int d^2x~\chi^T(\v x)(-i IZ~{\p_1}-i IX~{\p_2}-i\phi Z\e)\chi(\v x) &\mbox{if } x\geq0 \\
\int d^2x~\chi^T(\v x)(-i IZ~{\p_1}-i IX~{\p_2}-i\phi_3 I\e)\chi(\v x) &\mbox{if } x<0 \\
\end{cases}
\end{align*}
It is seen that $ ZI $ commutes with the hamiltonian and the Hilbert space is decomposed into two sectors with eigenvalues $ \pm1 $ under $ ZI $. The difference between the two sectors is that the sign of the $ \phi $ term changes. Without loss of generality we may assume $ \phi_3>0 $. Then the Hamiltonian in the $ ZI=-1 $ sector reads
\begin{align*}
H_{+}=&\begin{cases}
\int d^2x~\chi^T(\v x)(-i Z~{\p_1}-i X~{\p_2}+i\phi \e)\chi(\v x) &\mbox{if } x\geq0 \\
\int d^2x~\chi^T(\v x)(-i Z~{\p_1}-i X~{\p_2}-i\phi_3 \e)\chi(\v x) &\mbox{if } x<0 \\
\end{cases}
\end{align*}
An eigenmode $ \xi=\int d^2x~ f^{T}(\v x) \chi(\v x) $ satisfying $ \left[H_{+},\xi\right]=2E\xi $ implies $ Dg=E $, where
\begin{align*}
D=&\begin{cases}
-i Z~{\p_1}+ X~{k_2}+i\phi \e &\mbox{if } x\geq0 \\
-i Z~{\p_1}+ X~{k_2}-i\phi_3 \e &\mbox{if } x<0 \\
\end{cases}
\end{align*}
where we assumed $ f(x,y)=e^{ik_2 y} g(x) $ by translation invariance along $ \hat{y} $. The solution is
\begin{align*}
g(x)\propto&\begin{cases}
\exp(\int dx~Y(k_2I-EX)+\phi X) &\mbox{if } x\geq0 \\
\exp(\int dx~Y(k_2I-EX)-\phi_3 X) &\mbox{if } x<0 \\
\end{cases}
\end{align*}
which is normalizable in the $ X=-1 $ sector and when $ E=-k_2 $. All solutions in the other sectors are not normalizable or continuous at $ x=0 $. Thus there is only one chiral majorana mode in the interface.

\subsection{Numerical solution}
The corresponding lattice versions of \eqref{eq:scfm} are
\be
H_{SC}=\sum_i -\chi_{i}^{T} \left[(1+\phi_1)iX\e + (1+\phi_2) iZ\e \right]\chi_i
 + \sum_i \left[\chi_{i}^{T} \left(\frac{-i IZ+iX\e}{2}\right)\chi_{i+\hat{x}}+\chi_{i}^{T} \left(\frac{-i IX+iZ\e}{2}\right)\chi_{i+\hat{y}} + h.c.\right],\nn
 && \label{edgeHSC}\ee
 \be
H_{QAH}=\sum_i -\chi_{i}^{T} \left[iX\e +  iZ\e + \phi_3~iI\e\right]\chi_i
 + \sum_i \left[\chi_{i}^{T} \left(\frac{-i IZ+iX\e}{2}\right)\chi_{i+\hat{x}}+\chi_{i}^{T} \left(\frac{-i IX+iZ\e}{2}\right)\chi_{i+\hat{y}} + h.c.\right],\nn
 && \label{edgeHFM}\ee

We may perform an exact diagonalization of the single body Hamiltonian defined on a cylinder geometry (Fig \ref{fig5}(a)) where a ferromagnetic region is sandwiched between two superconducting regions. The periodic directions is $ \hat{x} $ and hence we can assume $ \chi(\v x)=e^{ik_x}\chi(y) $, so
\begin{align*}
H_{SC}=\sum_{i,k_x} \chi_{i}^{T} \left[\sin(k_x)IZ-(1-\cos(k_x)+\phi_1)iX\e - (1+\phi_2) iZ\e \right]\chi_i
 + \sum_i \left[\chi_{i}^{T} \left(\frac{-i IX+iZ\e}{2}\right)\chi_{i+\hat{y}} + h.c.\right]\\
 H_{QAH}=\sum_{i,k_x} \chi_{i}^{T} \left[\sin(k_x)IZ-(1-\cos(k_x))iX\e - iZ\e - \phi_3~iI\e \right]\chi_i
  + \sum_i \left[\chi_{i}^{T} \left(\frac{-i IX+iZ\e}{2}\right)\chi_{i+\hat{y}} + h.c.\right]
\end{align*}

We observe two counter propagating majorana modes in the single body spectrum. We plotted the corresponding wavefunctions and find that the left-moving/right-moving chiral majorana modes are respectively localized on the upper/lower interface between the superconducting and ferromagnetic regions. The result still holds when a chemical potential term $ \Delta H=-\mu\chi^T Q \chi $ with $ Q=YI $ is added to the Hamiltonians. See Fig \ref{fig5}.

\begin{figure}
\includegraphics[scale=0.35]{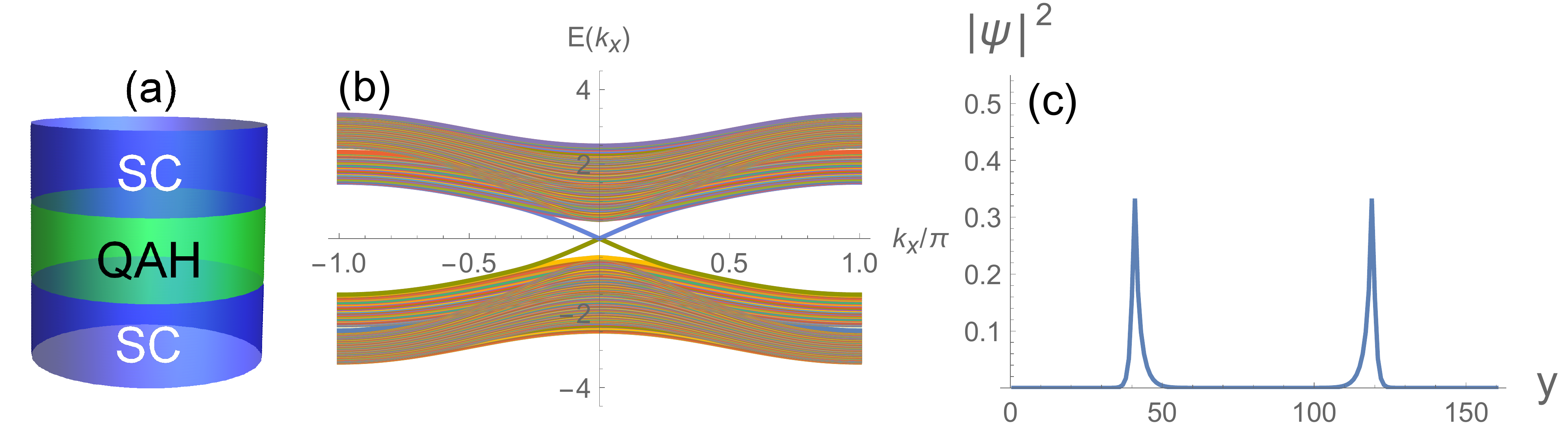}
\caption{(a) The interface between superconducting and ferromagnetic regions.(b) The projected bandstructure where $k_x$ is the momentum around the circumference of the cylinder. (c) The wavefunction of the in-gap state at $k_x=0.2 \pi$. The left and right Majorana modes are localized on the upper and lower inteferface, respectively.The parameters useds are $\phi_1=1.0\cos(\pi/8),\phi_2=1.0\sin(\pi/8),\phi_3=0.4$. The height of the cylinder is $160$ lattice spacings. The widths of the three different regions are $40-80-40$.}
\label{fig5}
\end{figure}
\subsection{The mass manifolds of the no symmetry class}
In \Fig{gapclosure}(b) we show the mass manifolds associated with the free fermion symmetry protected topological states in the no symmetry class. The superconducting mass manifold in  \Fig{gapclosure}(a) has $T^2=+1$ symmetry, however since it is
connected to that of \Fig{gapclosure}(b) without gap closing they describe the same topological phase.
\begin{figure}
\includegraphics[scale=0.35]{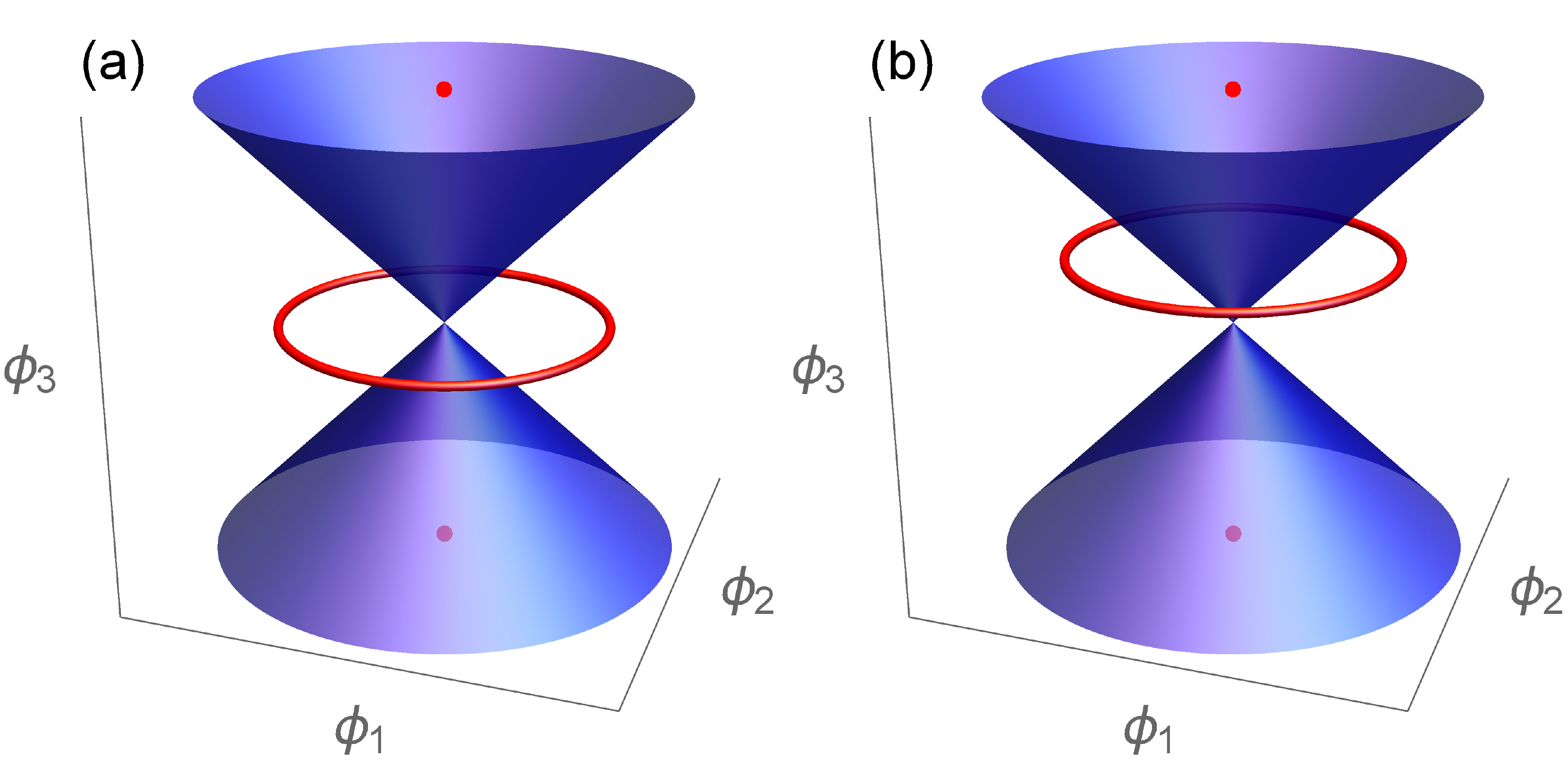}
\caption{(a) The mass manifold of the no symmetry class consists of two points (red) and a circle (red). The blue conic surface marks where the fermion gap closes. The hamiltonian associated with the circle is \Eq{H2d}. (b) By shifting the circle up the $T^2=+1$ symmetry of the superconductor is broken. However because the shifted circle does not cross the gap closing surface, the topological class of the phase does not change. }
\label{gapclosure}
\end{figure}
\end{widetext}

\begin{thebibliography}{20}
\bibitem{ReadGreen} N. Read and D. Green, Phys. Rev. B {\bf 61}, 20167 (2000).
\bibitem{Alicea} For a review see J. Alicea, Rep. Prog. Phys. {\bf 75}, 076501 (2012).
\bibitem{FuKane} L. Fu and C.L. Kane, Phys. Rev. Lett. {\bf 100}, 096407 (2008).
\bibitem{JFJia} H.H. Sun {\it et al.}, Phys. Rev. Lett. {\bf116}, 257003 (2016).
\bibitem{HDing} P. Zhang {\it et al.},  Science 08 Mar 2018, eaan4596.
\bibitem{Kitaev2009}A. Kitaev, AIP conference proceedings, {\bf 1134},20 (2009).
\bibitem{Wen2012} X.-G. Wen, Phys. Rev. B {\bf 85},085103 (2012).
\bibitem{Teo} J. Teo and C.L. Kane, Phys. Rev. B {\bf 82}, 115120 (2010).
\bibitem{Bott1959} R. Bott, Annals of Mathematics, {\bf 70}, 313 (1959).
\bibitem{yu} R. Yu {\it et al}, Science {\bf329}, 61 (2010).
\bibitem{xue} C.-Z. Chang {\it et al}, Science {\bf 340}, 167 (2013).
\bibitem{Qi}X.-L. Qi, T. L. Hughes and S.-C. Zhang, Phys. Rev. B {\bf82}, 184516 (2010).
\bibitem{JackiwRossi}R. Jackiw  and P. Rossi, Nucl. Phys. B{\bf 190}, 681 (1981).
\end{thebibliography}
\end{document}